\documentclass[twocolumn,aps,prb,floatfix,showpacs]{revtex4}

\usepackage{graphicx}

\begin{document}

\title{Probing Magnetic Gap Collapse within a Three-band Model of Resonant 
inelastic x-ray scattering (RIXS) in the Cuprates}
\author{R.S. Markiewicz and A. Bansil}

\affiliation{
%\address{
Physics Department, Northeastern University, Boston MA 02115, USA}

\begin{abstract}

We present a three-band Hubbard Hamiltonian and the associated Cu K-edge 
RIXS spectra for electron and hole doped cuprates over a wide range of 
energy and momentum transfers.  By comparing 
computed spectra for the unfilled case with the corresponding results for 
15\% electron or hole doping at two different values of the effective 
Hubbard parameter $U$, generic signatures of the collapse of the magnetic 
gap and the characteristic momentum dependencies and evolution of the 
spectra with doping are identified.  Available RIXS data support the gap collapse 
scenario for electron doped cuprates, but the situation in hole doped systems 
is found to be less clear.

\end{abstract}

%\pacs{PACS numbers~:~~74.20.Mn, 74.72.-h, 71.45.Lr, 74.50.+r }

\date{\today}

\maketitle

%****
%\narrowtext
%****

In a resonant inelastic x-ray scattering (RIXS) experiment, the incident 
x-ray photon is scattered after inducing electronic transitions in the 
vicinity of the Fermi energy ($E_F$), as in Raman scattering.\cite{KoShin} 
Since no charged particles enter or leave the 
sample, the technique is attractive as a genuine bulk probe 
and a new spectrocopic tool for getting a handle on 
the nature of both filled and unfilled electronic states within a few 
eV's of $E_F$. Although filled states are also accessible via 
angle-resolved photoemission spectroscopy (ARPES), the unique ability of 
RIXS to access unfilled states makes it ideal for probing how the magnetic 
gap in insulating half-filled cuprates evolves with doping.  The momentum-space 
evolution of this process and the possible collapse of the magnetic gap 
at optimal doping are fundamental pieces of information necessary for identifying 
different routes from the insulating to the superconducting state and the mechanism 
which underlies the remarkable phenomenon of high-temperature superconductivity.

This article discusses a three band Hubbard Hamiltonian and the associated 
Cu K-edge RIXS spectra for energy transfer $\omega$ extending to 8 eV and 
momentum transfer $\bf q$ along the $(\pi,\pi)\rightarrow 
\Gamma\rightarrow (\pi,0)$ symmetry line. Our purpose is to obtain a 
systematic scheme for interpreting spectral features and their evolution 
with electron and hole doping in the cuprates. The specific parameters are 
chosen to describe Nd$_{2-x}$Ce$_x$CuO$_4$ (NCCO), but the results are expected to be 
representative of the cuprates more generally. The electronic structure is treated at the 
Hartree-Fock level with variable $U$ to simulate the collapse of the magnetic gap. 
The solution at half-filling for $U$ = 7.2 eV is found to be antiferromagnetically 
(AFM) ordered and to display a `charge transfer' gap of $\approx$ 2 eV between the 
Cu d and O-p orbitals. At fixed $U$=7.2 eV, the effect of electron doping is to move 
most RIXS transitions to lower energies (in relation to half-filling), 
while hole doping introduces new `bands' of transitions in the spectrum, 
although both electron and hole doping yield additional weak transitions 
at low energy transfers in the 0-2 eV range. For $U$=5.7 eV, on the other 
hand, where the AFM gap collapses near optimal electron or hole doping, 
strong features in the RIXS spectra are predicted with characteristic 
${\bf q}$-dependencies at low energies. In confronting our predictions with 
available experimental 
results\cite{Abba,Hill1,HHH,Hasan2,Hill3,Hasan4,ITE,LuG}, we find that the 
RIXS data on NCCO support the scenario of a collapsed magnetic gap at 
optimal doping, while the situation with the hole doped systems is less 
clear. Our study suggests regions of $\omega-\bf q$ space where further 
experimental work would be valuable.

As general background, we note that Cu $K$-edge RIXS is an indirect probe 
of the magnetic gap, since the x-ray excites a $1s\rightarrow4p$ core 
electron, leading to a shake-up of the near-Fermi level states associated 
with screening the core hole.  The relative merits of indirect and direct 
RIXS in the cuprates have been debated\cite{UTI,ZC,VF}, where the latter 
refers to $L$-edge RIXS and the excitation of an electron from the Cu $2p$ 
directly into the Cu $3d$ states.  The polarization selection rules in 
RIXS are equivalent to those in Raman.  Screening excitations in K-edge 
RIXS satisfy $\Delta l$=0,2 selection rule, so that the RIXS gap can be 
different from the optical gap ($\Delta l$=1)\cite{ZC}.

Most treatments of RIXS spectra are based on cluster 
models\cite{Hill1,HHH,Hasan2,TTM2,IKo,Hasan3,ITE}, which are 
intrinsically limited in describing solid-state band dispersions, although 
some studies have involved single band, infinite dimensional 
models\cite{Dev1}. We consider a 3-band Hubbard model 
Hamiltonian with Cu $d_{x^2-y^2}$ and two O $p-\sigma$ orbitals:
\begin{eqnarray}
H=\sum_j(\Delta_0 d^{\dagger}_jd_j+Un_{j\uparrow}n_{j\downarrow})
+\sum_iU_pn_{i\uparrow}n_{i\downarrow}
\nonumber \\
+\sum_{<i,j>}t_{CuO}[d^{\dagger}_jp_i+(c.c.)] 
%\nonumber \\
+\sum_{<<i,i'>>}t_{OO}[p^{\dagger}_{i}p_{i'}+(c.c.)].
%\nonumber \\
\label{eq:100}
\end{eqnarray}
\begin{figure}
\begin{center}
%   \resizebox{8.5cm}{!}{\includegraphics{3brixs37.eps}}
   \resizebox{8.5cm}{!}{\includegraphics{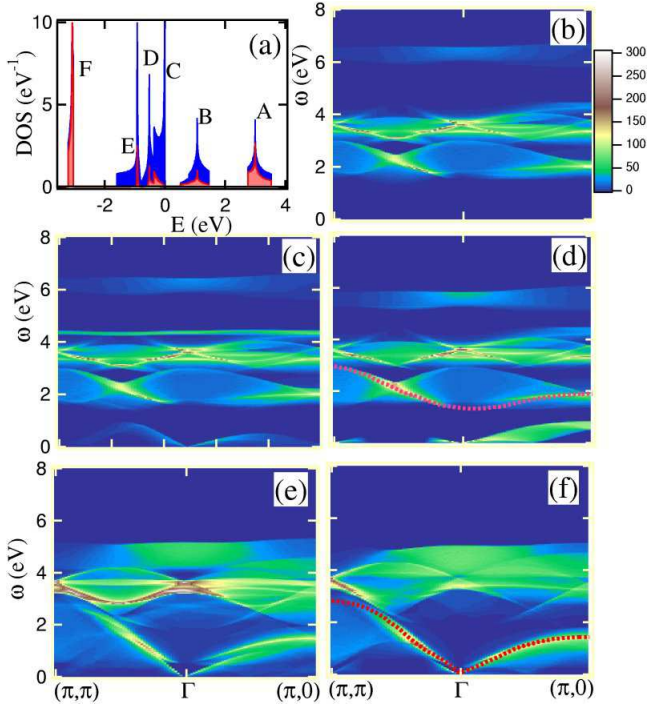}}
\end{center}
%\leavevmode
%   \epsfxsize=0.33\textwidth\epsfbox{3brixs9.eps}  
%   \epsfxsize=0.44\textwidth\epsfbox{3brixs14.eps}  
%   \epsfxsize=0.44\textwidth\epsfbox{3brixs31.eps}  
%   \epsfxsize=0.44\textwidth\epsfbox{3brixs33.eps} 
%\vskip0.5cm
\caption{\label{fig:4}
(color) 
(a) Density of states (DOS) and its Cu (red) and O (blue) components at 
half-filling for $U$ = 7.2 eV. (b-f) Maps of RIXS intensity spectra 
in the $\omega$-{\bf q} space approximated by WK-JDOS for: (b) half-filling; 
(c) 15\% holes, $U$ = 7.2 eV; (d) 15\% 
electrons, $U$ = 7.2 eV; (e) 15\% holes, $U$ = 5.7 eV; and (f) 
15\% electrons, $U$ = 5.7 eV.  Dotted lines in (d) and (f) highlight 
prominent low energy features.
}
\end{figure}

\noindent
Here $\Delta_0$ is the bare on-site energy of the Cu orbital with respect 
to that of the O orbitals (taken to define the energy zero), $t_{CuO}$ and 
$t_{OO}$ are the Cu-O and O-O nearest neighbor (NN) hopping parameters and 
$U$ [$U_p$] is the Cu [O] on-site Coulomb repulsion. The single [double] 
angular brackets on the summations denote that these sums are restricted 
to the Cu-O[O-O] NN terms. The specific values of the parameters have been 
obtained from an analysis\cite{foot1,MKII,Ar3} of the doping-dependent ARPES 
spectra of  NCCO: $t_{CuO}=0.8$ eV, $t_{OO}=-0.4$ 
eV, $\Delta_0 =-0.755$ eV, $U_p=5.0$ eV, and $U=7.2$ eV for $x=0$. $U$ is 
found to decrease with doping with a value of 5.7 eV at optimal doping, 
while the other parameters are doping independent. Accordingly, 
computations with $U$=5.7eV as well as $U$=7.2 eV are presented. Since similar 
parameter values describe LSCO\cite{Ar3}, we expect 
our computations to capture generic features of the RIXS spectra of 
both electron and hole doped cuprates.

As a prelude to RIXS computations, we solve the electronic structure 
at the level of the Hartree-Fock approximation. Note that the treatment of 
terms involving the Hubbard parameters $U$ and $U_p$, renormalize $\Delta_0$ 
to an effective parameter $\Delta =\Delta_0+Un_d/2-U_pn_p/2$, where $n_d$ 
and $m_d=(n_{d\uparrow}-n_{d\downarrow})/2$ are the average electron 
density and magnetization on Cu, with similar definitions for the oxygens. 
At half-filling, the self-consistent magnetization is found to be 
$m_d=0.295$ on Cu-sites, while the O-sites are unpolarized. The computed 
doping dependence of $m_d$ is similar in 3-band\cite{MKII} and one-band\cite{foot1} models.

For orientation, Fig.~\ref{fig:4}(a) shows the density of states (DOS) for 
$U$=7.2 eV at half-filling ($x=0$).  The AFM ordering yields a total of six 
bands marked A-F. At half filling $E_F$ lies in the gap between the 
two highest bands $-$ the `upper Hubbard band' A at $\sim 3$ eV and the 
`charge transfer' or `Zhang-Rice' band B at $\sim 1$ eV. Note that the 
splitting between bands A and F, which are dominated by the Cu 
contribution (red), is the main Mott gap controlled by $U$, while the A-B 
splitting is related to the product $Um_d$, which is the AFM gap\cite{foot5}.

The K-shell RIXS cross-section for the Cu $1s\rightarrow 4p$ core level 
excitation can be written as \cite{NoI,foot2}
\begin{eqnarray}
W({\bf q},\omega ,\omega_i)=(2\pi )^3N|w(\omega ,\omega_i)|^2
\nonumber \\
\times\sum_{{\bf k}jj'}\delta (\omega +E_j({\bf k})-E_{j'}({\bf k+q}))n_j({\bf 
k})[1-n_{j'}({\bf k+q})] 
\nonumber \\
\times |\sum_{\ell ,\sigma ,\sigma '}e^{i{\bf q\cdot R_{\ell}}}\alpha_{\ell}X^j_{\ell\sigma 
}\Lambda_{\sigma ,\sigma 
'}(\omega ,{\bf q})X^{j'}_{\ell\sigma '}|^2,
\label{eq:101}
\end{eqnarray}
where
\begin{eqnarray}
w(\omega ,\omega_i)=|\gamma|^2\sum_{\bf k_1}{V_d\over D(\omega_i,{\bf 
k_1})D(\omega_f,{\bf k_1})}.
\label{eq:102}
\end{eqnarray}
Here $D(\omega,{\bf k})=\omega+\epsilon_{1s}-\epsilon_{4p}({\bf 
k})+i\Gamma_{1s}$. The Cu $1s$ band is assumed dispersionless at energy 
$\epsilon_{1s}$, the Cu $4p$ band $\epsilon_{4p}({\bf k_1})$ is modelled 
by a 2D tight binding band with NN hopping, and the near-FS bands are 
described by energies $E_j({\bf k})$, occupation numbers $n_j({\bf k})$, 
and eigenvectors $X^j_{\ell\sigma}$ of spin $\sigma$ and orbital $\ell$ 
which include a magnetic structure factor. $\omega_i$ [$\omega_f$] and 
${\bf q_i}$ [${\bf q_f}$] are the energy and wave number of the incident 
[scattered] photon, with energy and momentum transfer given by, $\omega 
=\omega_i-\omega_f$ and ${\bf q=q_i-q_f}$.  Polarization effects arise via 
the Cu $1s-4p$ transition factor $\gamma$, which for simplicity is taken 
as a constant. The core hole broadening is treated phenomenologically by 
the damping parameter $\Gamma_{1s}$ with a value of 0.8 eV\cite{LuG,NoI}. 
$N$ is the total number of Cu atoms. $V_d$ is the effective potential for 
creating an on-site Cu $3d$ electron-hole pair. Although the predominant 
excitations are expected to be on-site Cu, we include the possibility of 
NN O excitations via $\alpha_1=V_1/V_d$, where $V_1$ is the potential for 
O $2p$ excitations, as well as second NN Cu excitations via 
$\alpha_{2}=V_{2}/V_d$.  
%The factor of $\alpha_d\equiv 1$ is introduced 
%for convenience in Eq. 2.  
We assume small values of $\alpha_1=0.1$, 
$\alpha_2=0.05$ in this study.  ${\bf R_{\ell}}$ is a vector from the 
core hole to the atom on which the electron-hole pair is excited. Finally, 
$\Lambda_{\sigma ,\sigma '}$ is a vertex correction describing excitonic 
coupling of the electron-hole pair; we set 
$\Lambda_{\sigma,\sigma'}=\delta_{\sigma,\sigma'}$, which is generally a 
good approximation\cite{NoI}.

If the matrix element (ME) $w(\omega ,\omega_i)$ of Eq.~\ref{eq:102} is 
approximated as a constant, then Eq.~\ref{eq:101} becomes a relatively 
simple weighted {\bf k}-resolved joint density of states (WK-JDOS) involving 
bands $j$ and $j'$ where the weights involve only Cu$\rightarrow$Cu or 
O$\rightarrow$O transitions. More generally, at a fixed photon energy 
$\omega_i$, the factor $|w(\omega ,\omega_i)|^2$ provides an 
$\omega$-dependent modulation of the WK-JDOS. Computations with and 
without the RIXS cross-section approximated as WK-JDOS will be considered.

Fig. 1 shows the RIXS cross-section approximated as WK-JDOS for a number 
of different cases to allow the unfolding of effects of electron or hole 
doping and those due to the collapse of the AFM gap. We focus first on the 
RIXS `bands'\cite{foot3} in (b) for the undoped system and their 
relationship to the six DOS `bands' A-F in (a). Here each of the five 
filled bands B-F can be excited into the unfilled band A, yielding five 
RIXS bands in (b). The lowest RIXS band comes from B $\rightarrow$ A 
transitions and lies around 2 eV, while the highest band around 6 eV in 
(b) comes from F $\rightarrow$ A transitions. The overlapping complex of 
bands around 3.5 eV is related to transitions from the O-related DOS bands 
C-E. With adequate resolution and statistics, RIXS should be able to map 
the RIXS bands of the sort shown in (b) and thus adduce the nature of the 
near-$E_F$ electronic structure in considerable detail.

We next consider the effect of electron doping by comparing (b) and (d), 
where (d) refers to 15\% electron doping at $U$=7.2 eV. Electron doping 
adds electrons to the bottom of the DOS band A, causing the portion of 
band A available for transitions to become narrower, so that transitions 
from bands B-E $\rightarrow$ A will display a smaller spread in energy. At 
a fixed value of $m_d$ this will induce a shift in the RIXS bands to 
higher energies, but this effect is counteracted by a reduction of $m_d$ 
with doping (even when $U$ is held fixed). For example, the B 
$\rightarrow$ A RIXS band near $\Gamma$ in (d) finally lies at a lower 
energy of $\approx$ 0.2 eV than in (b). Also, the added electrons at the 
bottom of A can be excited intraband within A and yield a new RIXS band at 
low energies below 1 eV in (d), especially along the $\Gamma-M(\pi,0)$ 
line.

The effect of hole doping is fundamentally different in that holes remove 
states from the top of band B. Since bands A and C-F remain unchanged, 
transitions from C-F $\rightarrow$ A will see little change, but the 
narrowing of band B due to added holes would move B $\rightarrow$ A 
transitions to higher energies (if $m_d$ remained constant).\cite{foot4}  Moreover, 
the added holes allow the formation of four new RIXS bands associated with 
excitations from C-F to the top of B as seen, for example, by comparing 
(b) and (c) at the middle of the $\Gamma-M$ line around 1.4 eV and at 
$\approx$ 4.3 eV. New intraband transitions within B also become possible, 
and much like the case of electron doping discussed above, these appear at 
low energies below 1 eV. However, the {\bf k}-dependencies of such low 
energy RIXS bands for the hole and electron doping are seen by comparing 
(c) and (d) to differ substantially, reflecting the differing {\bf k}-space locations of 
the respective band edges.

While $m_d$ decreases with doping at fixed $U$, the resulting magnetic gap 
is still too large to explain the ARPES spectra of NCCO, suggesting that 
the mean field theory does not properly account for screening of $U$.  
Reducing $U$ to 5.7 eV leads to a near collapse of the band splitting with 
$m_d$ = 0.063, in good agreement with NCCO ARPES data. [For hole doping, 
$U$=5.7 eV yields $m_d$=0.12.] The effects on the RIXS spectrum can be 
seen by comparing Fig. 1(d) with (f) for electron doping and Fig. 1(c) 
with (e) for hole doping. The most prominent signature of the collapse of 
the AFM gap is that the lowest RIXS band is shifted down to nearly zero 
energy at $\Gamma$.

\begin{figure}
\begin{center}
%   \resizebox{8.5cm}{!}{\includegraphics{3brixs210.eps}}
   \resizebox{8.5cm}{!}{\includegraphics{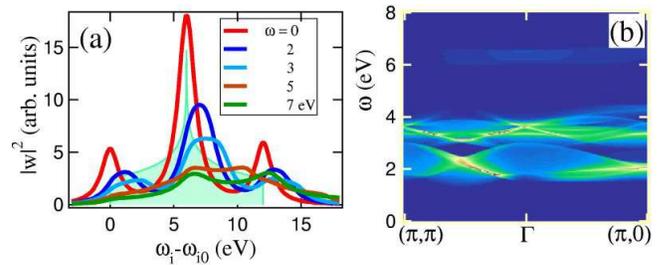}}
\end{center}
%   \epsfxsize=0.33\textwidth\epsfbox{3brixs8.eps}  
%   \epsfxsize=0.46\textwidth\epsfbox{3brixs27.eps}  
%   \epsfxsize=0.46\textwidth\epsfbox{3brixs29.eps}  
%\leavevmode
%\vskip0.5cm
\caption{\label{fig:2}
(color) 
(a) RIXS intensity factor $|w|^2$ as a function of $\omega_i$ for several 
values of $\omega$. $\omega_{i0}$ is the 1$s$-4$p$ threshold energy. Green 
shaded area schematically shows the DOS for the Cu $4p$ band. (b) RIXS 
intensity map at $\omega_i=\omega_{i0}$ for half-filling at $U$=7.2 eV.}
\end{figure}

Fig.~\ref{fig:2} illustrates the nature of the factor $w(\omega,\omega_i)$ 
of Eq.~3 and its effect on the spectra. (a) shows the behavior of $|w|^2$ 
as a function of $\omega_i$ for several values of $\omega$. [Note that 
Eq.~3 for $w$ does not depend on ${\bf q}$.] The peaks in $|w|^2$ are seen 
to be tied to the Van Hove singularities in the underlying 
$4p$ DOS (green shaded area).  In (b), the full RIXS spectrum of Eqs. 1 
and 2 for $x=0$, including the effect of $w$, is shown for a particular 
photon energy. A comparison of Figs. 2(b) and 1(b) indicates 
that the modulation of the spectra due to $|w|^2$ is relatively weak.

Interestingly, the AFM order shows up in the $U$=7.2eV spectra in 
Figs.~1(b)-(d) and Fig.~2(b) as an approximate symmetry in the RIXS bands 
about the $(\pi/2,\pi/2)$ point along the M-$\Gamma$ line. However, for 
$U$=5.7eV, the reduced magnetism leads to a loss of this symmetry as seen 
from Figs. 1(e) and (f). Note that the underlying bands in all cases possess 
the symmetry of the AFM order\cite{foot5}, but that this symmetry is in 
general not obeyed by various spectroscopic measurements [e.g., the 
`shadow bands' seen in ARPES].

\begin{figure}
\begin{center}
%   \resizebox{8.5cm}{!}{\rotatebox{-90}{\includegraphics{Fig3.eps}}}   
%   \resizebox{8.5cm}{!}{\rotatebox{-90}{\includegraphics{Fig3_new_dec2005_v2.eps}}}   
   \resizebox{8.5cm}{!}{\rotatebox{-90}{\includegraphics{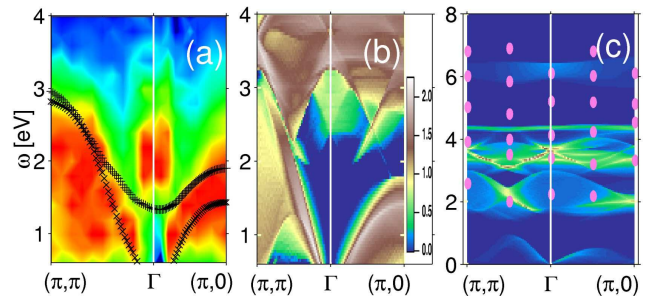}}}   
\end{center}
%\leavevmode
%   \epsfxsize=0.33\textwidth\epsfbox{3brixs12.eps}  
%   \epsfxsize=0.46\textwidth\epsfbox{3brixs15.eps}  
%   \epsfxsize=0.46\textwidth\epsfbox{3brixs32.eps}  
%   \epsfxsize=0.46\textwidth\epsfbox{3brixs34.eps}  
%\vskip0.5cm
\caption{\label{fig:1}
(color) 
(a) Experimental NCCO RIXS data from Ref. \protect\onlinecite{ITE}. 
Prominent low energy features marked by dotted lines in Figs. 1(d) and (f) 
for 15\% electron doping are reproduced for $U$ = 7.2 eV ($+$'s) or 5.7 eV 
($\times$'s). (b) Spectrum of Fig. 1(f) on a logarithmic scale to 
highlight low intensity features. (c) Experimental RIXS features in Hg1201 
(red dots) from Ref.~\protect\onlinecite{LuG} superposed on the RIXS map 
of Fig. 1(c) for 15\% holes at $U$ = 7.2 eV.}
\end{figure}

We comment now on the implications of some of the available experimental 
RIXS data in the light of our computations. In this connection, the 
RIXS data of Ref.~\onlinecite{ITE} from optimally doped NCCO is reproduced 
in Fig. 3(a), where the lowest RIXS bands highlighted in Figs. 1(d) and 
1(f) are superposed as $+$'s and $\times$'s, respectively. In comparing 
the theoretical spectra of Figs. 1(d) and 1(f) in the low energy regime of 
up to $\approx$ 3 eV with the results of Fig. 3(a), it is clear that the 
experiments are in substantially better accord with the results of Fig. 
1(f) than 1(d).  In particular, the data of Fig. 3(a) display a 
larger dispersion and presence of intensity extending to quite low 
energies along $(\pi,\pi)\rightarrow\Gamma$, while along the $\Gamma\rightarrow
(\pi,0)$ line the data show a smaller dispersion and are concentrated in 
a narrower window of energies, rather like the results of Fig. 1(f). In 
sharp contrast, Fig. 1(d) shows little intensity in the 0-1.5 eV range. 
Since the computations of Fig. 1(f) for $U$=5.7 eV involve a collapsed AFM 
gap, while those of Fig. 1(d) for $U$=7.2 eV do not, we are lead to 
adduce that the RIXS data on NCCO support an AFM gap collapse scenario. 

Fig. 3(a) also shows a feature around 2 eV at $\Gamma$.  The intensity of 
this feature, assigned to interband transitions, is enhanced in the 
representation of Fig. 3(a) where data are normalized to maximum at each 
$k$-point after subtracting high energy contributions. In reasonable accord, 
a weak feature around 2.5 eV at $\Gamma$ is seen in our computations in Fig. 
3(b), which is due to interband transitions from the top of the C-E band 
complex to the upper magnetic band\cite{figplot,otoc}. However, we 
should keep in mind that a quantitative description of the RIXS spectra 
with increasing energy transfer will require the inclusion of additional 
Cu-O and O bands beyond the present three-band model.  
For this reason, other high energy features present in the spectra of Ref.~\onlinecite{ITE} 
and in the theoretical spectra of Fig. 3(b) are not analyzed further here.

%At higher energies, we expect two additional bands, associated with 
%transitions from the bonding and nonbonding bands to the unoccupied 
%states 
%of the antibonding band.  A bonding band feature was found in 
%Ref.~\onlinecite{ITE}, but has been subtracted from the data of Fig. 3(a) 
%to enhance the lower energy features. In particular, there is an 
%additional feature near $\Gamma$ at about 2eV, in reasonable accord with 
%a 
%weak transition in Fig. 1(f) at $\sim 2.5$eV near $\Gamma$, due to 
%transitions from the top of the nonbonding band (corresponding to bands 
%C-D in Fig. 1(a)).  While the feature is weak, it can be made to resemble 
%Fig. 3(a) by replotting it in a similar fashion\cite{figplot}. 
%However, we should keep in mind that a satisfactory description of the 
%RIXS spectra with increasing energy transfer will require the inclusion 
%of 
%a number of bands beyond the present three-band model.

%However, our computations also make predictions for the high-energy 
%behavior of the RIXS spectra. By comparing Figs. 1(d) and (f) we see 
%that the collapse of the AFM gap induces a movement of RIXS bands at 
%highest energies to lower energies. Such a `compression' of the RIXS 
%spectra has so far not been observed clearly and further doping dependent 
%high energy experiments would be worthwhile. 

The situation in hole-doped cuprates is less clear.  
Earlier RIXS experiments\cite{Hill3,Hasan4} reported a few, strongly 
dispersing peaks, but recently Lu, et al.,\cite{LuG} find several 
additional weakly dispersing bands both in optimally doped 
HgBa$_2$CuO$_{\bf 4+\delta}$ (Hg1201) and in a reanalysis of the LSCO 
data. Fig. 3(b) shows the measured peak positions in Hg1201 overlayed on 
the computed RIXS spectrum of Fig. 1(c) for $U$=7.2eV. The placement in 
energy of the observed peaks (red dots) is in qualitative accord, but it 
is difficult to make any definitive assigments of spectral features in 
view of the rather sparse set of ${\bf q}$-values sampled in the experiment. 
With reference to Fig. 1(e) for small $U$, the most discriminating test of 
AFM gap collapse is the position of the prominent dispersing feature at low energies  
along the $\Gamma\rightarrow (\pi /2,\pi /2)$ line.  While a recent 
STM study\cite{Seam} hints that the collapse of the AFM gap is far from complete in 
at least some compounds, optical spectra of hole-doped 
cuprates find large changes in the lowest gap with doping\cite{Uch}.  Some differences 
between the optical and RIXS spectra may arise due to presence of the 
core hole in the latter case.  Nanoscale phase separation may also play a role. 
The appearance of new spectral features with 
hole doping (e.g. at 4.3 eV and at 1.4 eV at $(\pi/2,\pi/2)$ in Fig. 
1(d)), delineated above in connection with the discussion of Fig. 1, would 
be tell-tale signatures of the uniformly doped state.  

%hole-doped cuprates the hole distribution is probably spatially 
%nonuniform, associated with `stripe' physics. Moreover, due to the core 
%hole, screening may be incomplete\cite{PlI}, leading to a larger (less 
%screened) value of $U$. In a stripe picture the core hole should repel 
%the charged stripes, much as in the `Swiss cheese' model of Zn 
%doping\cite{Uem}.  Indeed, in the $Z+1$ picture\cite{PlI}, the Cu atom 
%with a core hole is {\it approximated by a Zn impurity}.

In conclusion, our study provides a systematic scheme for interpreting 
features in the RIXS spectra of the electron as well as hole doped 
cuprates, although the issue of nanoscale phase separation needs to be 
addressed in the latter case. 
In this way, signatures of collapse of the AFM gap in the RIXS 
spectra can be identified and their evolution with doping and momentum 
transfer vector at low energies can be connected to the behavior of the 
spectra at high energies to develop a robust understanding of the filled 
and unfilled states within a few eV's around the Fermi energy.

\begin{acknowledgments}

     This work is supported by the US Department of Energy contract
     DE-AC03-76SF00098 and benefited from the allocation of
     supercomputer time at NERSC and Northeastern University's Advanced
     Scientific Computation Center (ASCC).

\end{acknowledgments}


\begin{thebibliography}{99}
\bibitem{foot5}One-band model results [Ref.~\onlinecite{MKII}] suggest that the AFM gap
persists in the presence of short-range AFM order.
\bibitem{foot2}The present calculation differs from that of Ref.\onlinecite{NoI} by 
inclusion of $\alpha_i$, $i=1,2$ and a doping dependent $U$.
\bibitem{foot3}By RIXS `bands' we mean features in RIXS 
spectra such as those of Figs. 1(b)-(f) which arise from a particular pair 
of bands and present a band-like appearance. These should be distinguished 
from the conventional energy bands and the related DOS bands in Fig. 1(a).
\bibitem{foot4}In a one-band model only the lowest JDOS band is reproduced and electron- 
and hole-doping are nearly indistinguishable. [Ref.~\onlinecite{TTM2}]
\bibitem{figplot}
A plot of the spectrum of Fig. 3(b) over 0.5-3.0 eV by 
renormalizing to the maximum at each $q$-point, yields a theoretical map 
quite similar to that of Fig. 3(a) in highlighting the 2-3 eV region 
around $\Gamma$.
\bibitem{otoc}
The top of the C-E band complex in the present 3-band model mainly 
possesses O-like character. The weak intensity of the 2.5 eV feature in 
our computations reflects the limitation of the 3-band model in properly 
describing the Cu-character of states at the top of the C-E band complex.
\end{thebibliography}
\end{document}